\documentclass[]{emulateapj}

\usepackage{hyperref}
\usepackage{graphicx}

\usepackage{amsmath,amssymb}
\usepackage{bm}

\usepackage{natbib}
\usepackage{wasysym}

\shorttitle{Hydromagnetic instabilities in Neutron Stars}
\shortauthors{Lasky et al.}

\begin{document}

\title{Hydromagnetic instabilities in relativistic neutron stars}

\author{Paul D. Lasky, Burkhard Zink, Kostas D. Kokkotas and Kostas Glampedakis}
	\email{lasky@tat.physik.uni-tuebingen.de}
	\affil{Theoretical Astrophysics, IAAT, Eberhard Karls University of T\"ubingen, T\"ubingen 72076, Germany}

		\begin{abstract}
			We model the non-linear ideal magnetohydrodynamics of poloidal magnetic fields in neutron stars in general relativity assuming a polytropic equation of state.  We identify familiar hydromagnetic modes, in particular the 'sausage/varicose' mode and 'kink' instability inherent to poloidal magnetic fields.  The evolution is dominated by the kink instability, which causes a cataclysmic reconfiguration of the magnetic field.  The system subsequently evolves to new, non-axisymmetric, quasi-equilibrium end-states.  The existence of this branch of stable quasi-equilibria may have consequences for magnetar physics, including flare generation mechanisms and interpretations of quasi-periodic oscillations.
		\end{abstract}
		
		\received{6$^{\rm th}$ May 2011}
		\revised{20$^{\rm th}$ May 2011}
		\accepted{24$^{\rm th}$ May 2011}
		
		
		

\section{Introduction}
Neutron stars harbour the strongest known magnetic fields in Nature.  It is therefore not  surprising that the magnetic field plays a key role in many aspects of neutron star physics. Most of what is known about neutron star magnetism comes from the inferred dipole moment associated with the electromagnetic stellar spin-down. In systems with prominent thermal emission, the surface magnetic field strength can be independently measured by its influence on the emission pattern. Remarkably, various astronomical observations suggest a systematic difference in field strengths between different classes of neutron stars  \citep[see][for a recent review]{kaspi10}.  Aged systems, such as millisecond pulsars and accreting neutron stars in low-mass X-ray binaries, have relatively weak magnetic fields, $B\sim10^{8}\,\mbox{G}$.  Typical radio pulsars have $B\sim10^{12}\,\mbox{G}$.  The highest field strengths are associated with soft-gamma-repeaters (SGRs) and anomalous X-ray pulsars (AXPs) with magnetic fields $B\sim 10^{15}\,\mbox{G}$.

This last class of neutron stars, collectively known as magnetars, are the focus of this paper. These slowly-spinning objects exhibit high energy emission, which is occasionaly punctuated by bursts. In some rare occasions, powerful flares have been observed in SGRs. According to the prevailing magnetar model, as first put forward by \citet{duncan92,thompson95}, the energy reservoir powering magnetar activity is the energy of the magnetic field itself. The unusually high magnetar surface temperatures are likely the result of heating produced by the Ohmic decay of the magnetic field in the neutron star crust \cite[e.g.][]{pons07}. More spectacularly, giant SGR flares could be the manifestation of global magnetic field instabilities and of subsequent field reconfiguration \citep{thompson95}.  Furthermore, there is strong evidence that quasi-periodic oscillations (QPOs) observed in the light curves of these giant flares are of magneto-elastic nature (i.e. the restoring force is provided by magnetic and crustal elastic stresses) \citep[e.g.][]{levin06,levin07,sotani08a,cerdaduran09,gabler11}. Unfortunately, a detailed understanding of magnetar magnetohydrodynamics is hindered by the lack of information about the properties of the interior magnetic field. Such an understanding is a vital element if we are to carry out, for example, magnetar `asteroseismology' using the observed QPOs.  

The aforementioned reasons imply it is of great theoretical and astrophysical interest to model the dynamics of magnetic fields in the interior of magnetars (and in general of neutron stars). More specifically, three key questions have attracted the most attention to date: (i) What is the nature of long-term magnetic field equilibria in neutron stars? (ii) Do hydromagnetic instabilies play a role in phenomena like the bursts and flares seen in AXPs and SGRs? (iii) Which physical mechanism(s) determines the long-term evolution and decay of magnetic fields in neutron stars?

In this letter we make contact with these first two questions.  We study short-term (i.e. dynamical timescales) neutron star magnetohydrodynamics (MHD), specifically addressing the issue of global MHD instabilities in neutron stars and resulting magnetic field equilibria.  The last few years have seen a flare of activity on this subject, mainly motivated by magnetar astrophysics. A large body of work has been devoted to semi-analytical studies of axisymmetric MHD equilibria \citep[e.g.][]{ioka01,yoshida06a,yoshida06b,haskell08,ciolfi09,ciolfi10}, building on much earlier work regarding equilibria \citep[e.g.][]{monaghan65,roxburgh66,parker66} and stability analyses \citep{kruskal54,tayler57,tayler73,wright73,markey73,markey74,flowers77}. It is only recently that global, fully non-linear, MHD numerical calculations have appeared studying magnetic field stability and equilibria in stellar models \citep{geppert06,braithwaite06,braithwaite06b,braithwaite07,braithwaite08,braithwaite09,duez10,kiuchi11}.  The work presented herein falls into this last category.  We study the dynamical evolution of purely poloidal magnetic fields in neutron stars in general relativity, investigating intrinsic instabilities and subsequent equilibria, uncovering for the first time non-axisymmetric magnetic field equilibria in barotropic stars.

\section{Numerical Model}\label{numerical}
We study magnetized neutron stars through time evolution of the ideal MHD equations in general relativity under the Cowling approximation.  To this end, we utilise the three-dimensional GRMHD code {\sc thor} \citep{zink08,korobkin10} and her sister GPU code {\sc horizon} \citep{zink11}.  

The MHD portion of the code utilizes the conservative formalism outlined in \citet{anton06}.  Hyperbolic divergence cleaning is employed according to \citet{anderson06}.  We treat the exterior of the star as a low-density artificial atmosphere, allowing full evolution of the magnetic field in this region.  The outer boundary of our domain, located at approximately 1.4 times the stellar radius, uses Dirichlet boundary conditions for the magnetic field.  We have tested our evolutions against this condition and found no discernible difference with the nature of the instability or the end-state of the magnetic field configuration.  

We employ a Cartesian grid with $120^{3}$ grid points at single precision.  We have completed numerous resolution studies including $90^{3}$, $150^{3}$ and $200^{3}$ grid points, finding consistent phenomenology with converging growth-times for the instability.  Moreover, we have tested the use of double precision on these simulations and again found consistent phenomenology and timescales.

\subsection{Initial Conditions}\label{ICs}
We employ the spectral code {\sc lorene}\footnote{http://www.lorene.obspm.fr/}, which produces self-consistent solutions of the Einstein-Maxwell field equations in ideal MHD \citep{bocquet95}.  The spectral grid is mapped to our Cartesian grid and the system is allowed to evolve.  No additional perturbations are imposed, relying instead on intrinsic numerical noise.

\subsection{Mode Decomposition}
The modal structure of the instability is extracted by performing a Fourier decomposition of various physical quantities on a ring in the equatorial plane.  We compute complex weighted averages \citep{zink07}
\begin{align}
	C_{m}(f)=\frac{1}{2\pi}\int_{0}^{2\pi}f\left(\varpi,\phi,z=0\right){\rm e}^{im\phi}d\phi,\label{Cm}
\end{align}
where $\varpi=\sqrt{x^{2}+y^{2}}={\rm const.}$ lies in the initial equatorial plane of the magnetic field.  In equation (\ref{Cm}), $f$ is some quantity in the star chosen to best represent the instability.  Throughout much of this article we use the $\phi$ and $z$ components of the magnetic field, although note the velocity and density fields also exhibit the instability, however the signal is generally not as clean.

\subsection{Fiducial Neutron Star Model}\label{fiducial} 
Our fiducial model is a non-rotating star with polytropic equation of state relating pressure, $P$, and rest-mass density, $\rho$, through $P=K\rho^{\Gamma}$, with $\Gamma=2$ and $K=100$.  This gives a stellar model with gravitational mass $M=1.3\,\mbox{M}_{\odot}$ and equatorial radius $R=12.6\,\mbox{km}$.  Our fiducial model has an average field strength inside the star of $\bar{B}=1.3\times10^{16}\,\mbox{G}$, yielding an Alfv\'en crossing time of $t_{A}=5.0\,\mbox{ms}$.  Such models have polar magnetic field strengths of $B_{{\rm surf}}=8.8\times10^{15}\,\mbox{G}$, which is a factor of a few larger than observations of the dipole component of the strongest magnetic fields.  This is predominantly done to reduce computational costs. 

\section{Results}

\begin{figure}
\epsscale{1.0}
	\plotone{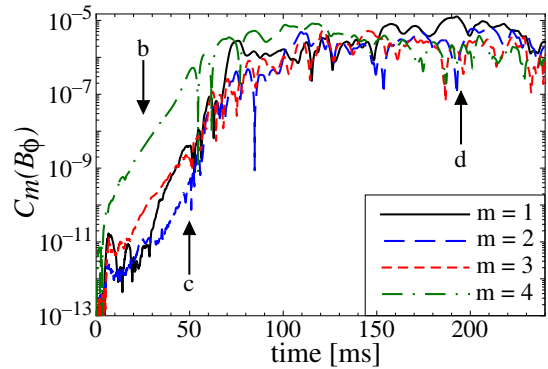}
	\caption{\label{fig1}  Evolution of $C_{m}(B_{\phi})$ as a function of time.  Our fiducial model has an average magnetic field of $\bar{B}=1.3\times10^{16}\,\mbox{G}$ implying an Alfv\'en crossing time of $5.0\,\mbox{ms}$.  The arrows represent the times of the three-dimensional snapshots plotted in figures \ref{fig2}.}
\end{figure}

In figure \ref{fig1} we plot the evolution of $m=1,\ldots,4$ modes for $C_{m}\left(B_{\phi}\right)$ for our fiducial model, measured at $\varpi=0.6\varpi_{\star}$, where $\varpi_{\star}$ is the stellar radius.  An instability is present in all modes after the first couple of Alfv\'en crossing times.  Each mode grows exponentially by approximately six orders of magnitude over many subsequent Alfv\'en crossing times.  Saturation of the modes occurs after about $75\,\mbox{ms}$, at which point the simulation evolves to a pseudo-equillibrium state.  We have evolved such simulations to $400\,\mbox{ms}=80\,t_{A}$, with little variation in the equilibrium configuration following the first hundred or so milliseconds. We discuss the equilibrium state in more detail below.  

In the first few milliseconds of the simulation we see the $m=4$ mode in figure \ref{fig1} grow by an order of magnitude over the other modes.  This feature is somewhat aligned with our Cartesian grid, which we attribute to a finite differencing effect associated with the surface of the star.  To understand this further we performed resolution studies with $150^{3}$ and $200^{3}$ grid points.  In such simulations we see the initial transient that lifts the $m=4$ mode above other modes slowly converging away with increasing resolution.  

\begin{center}
\begin{figure}[h]
\includegraphics[width=0.49\columnwidth]{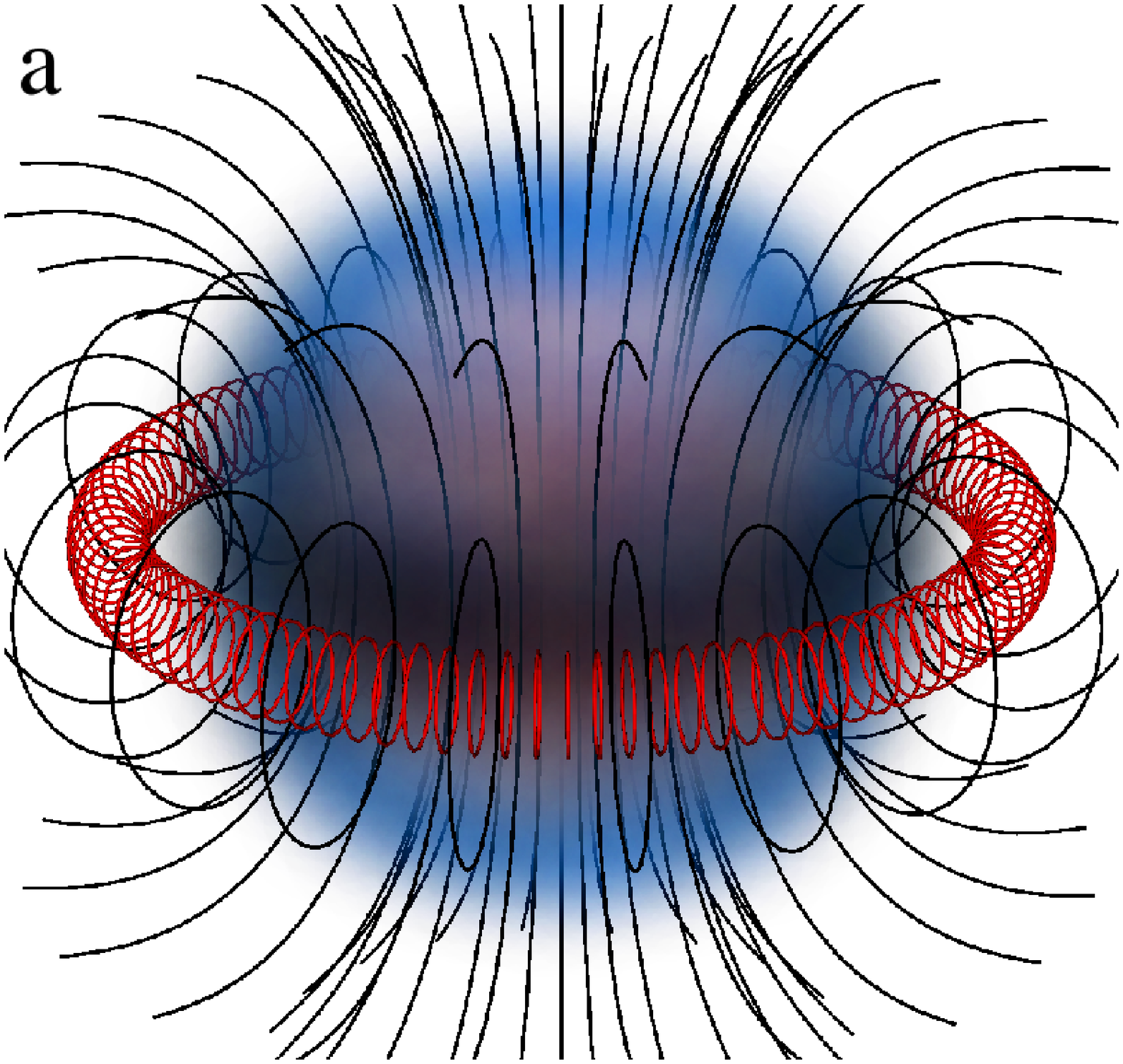}
\includegraphics[width=0.49\columnwidth]{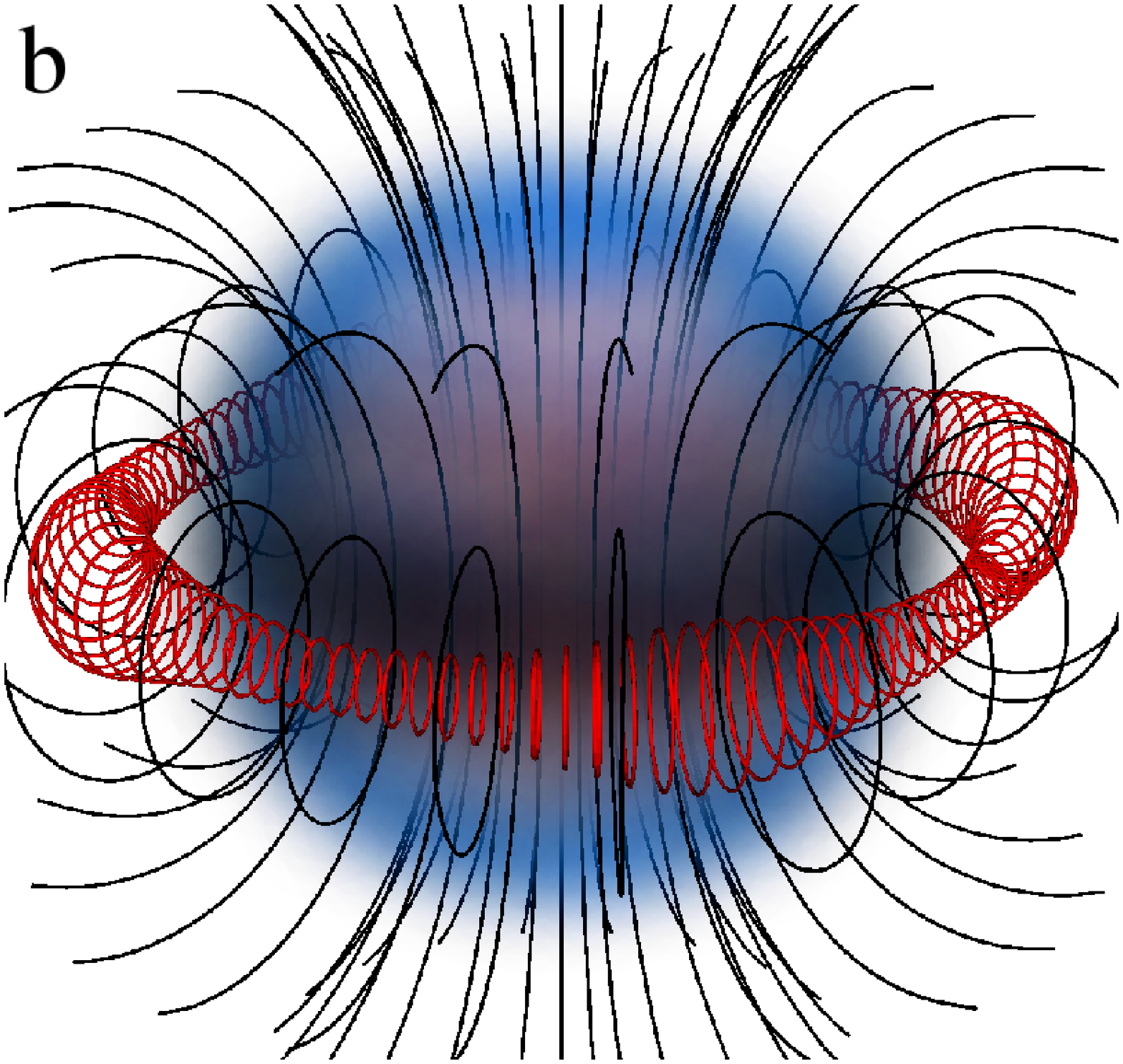}
\includegraphics[width=0.49\columnwidth]{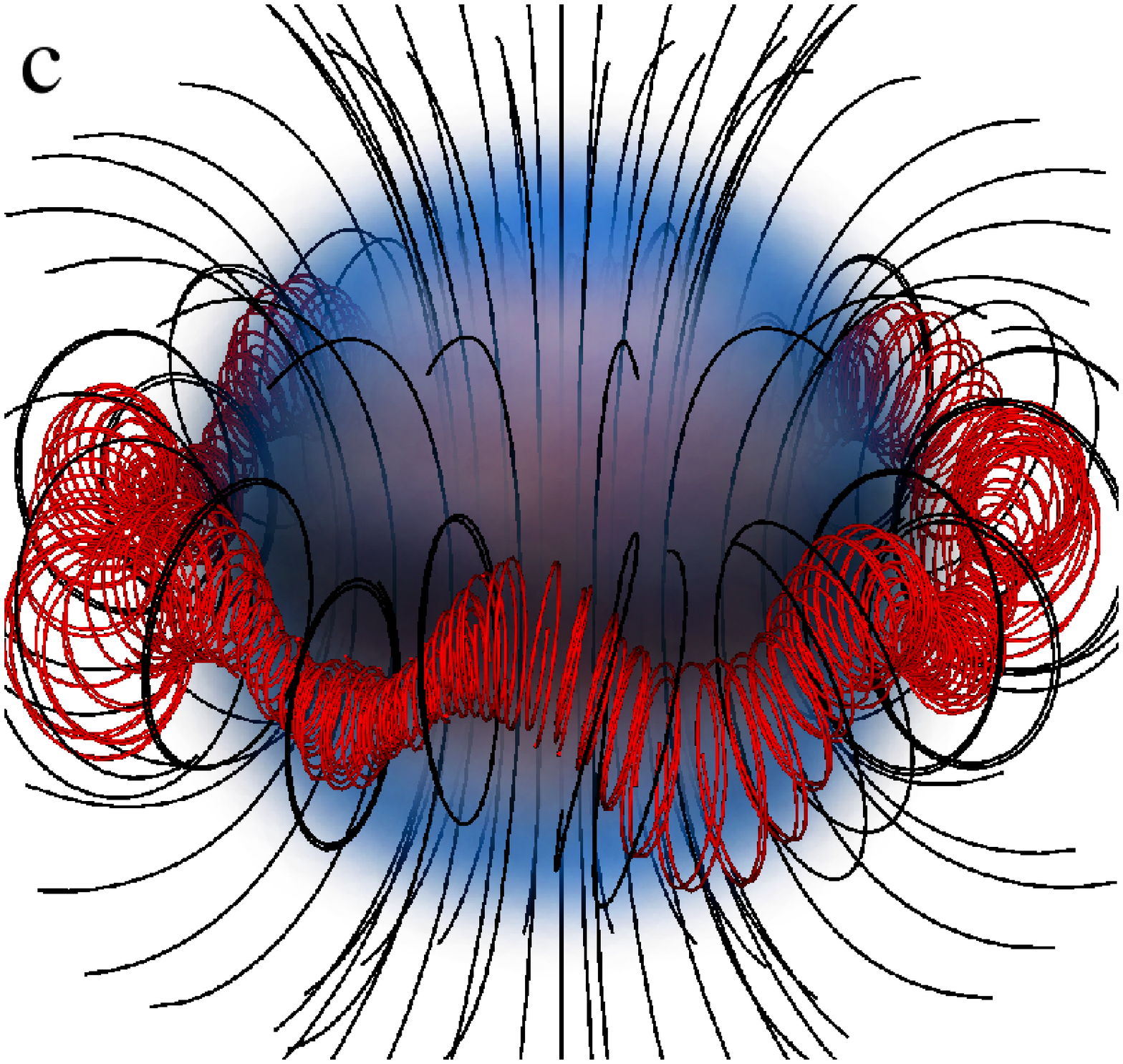}
\includegraphics[width=0.49\columnwidth]{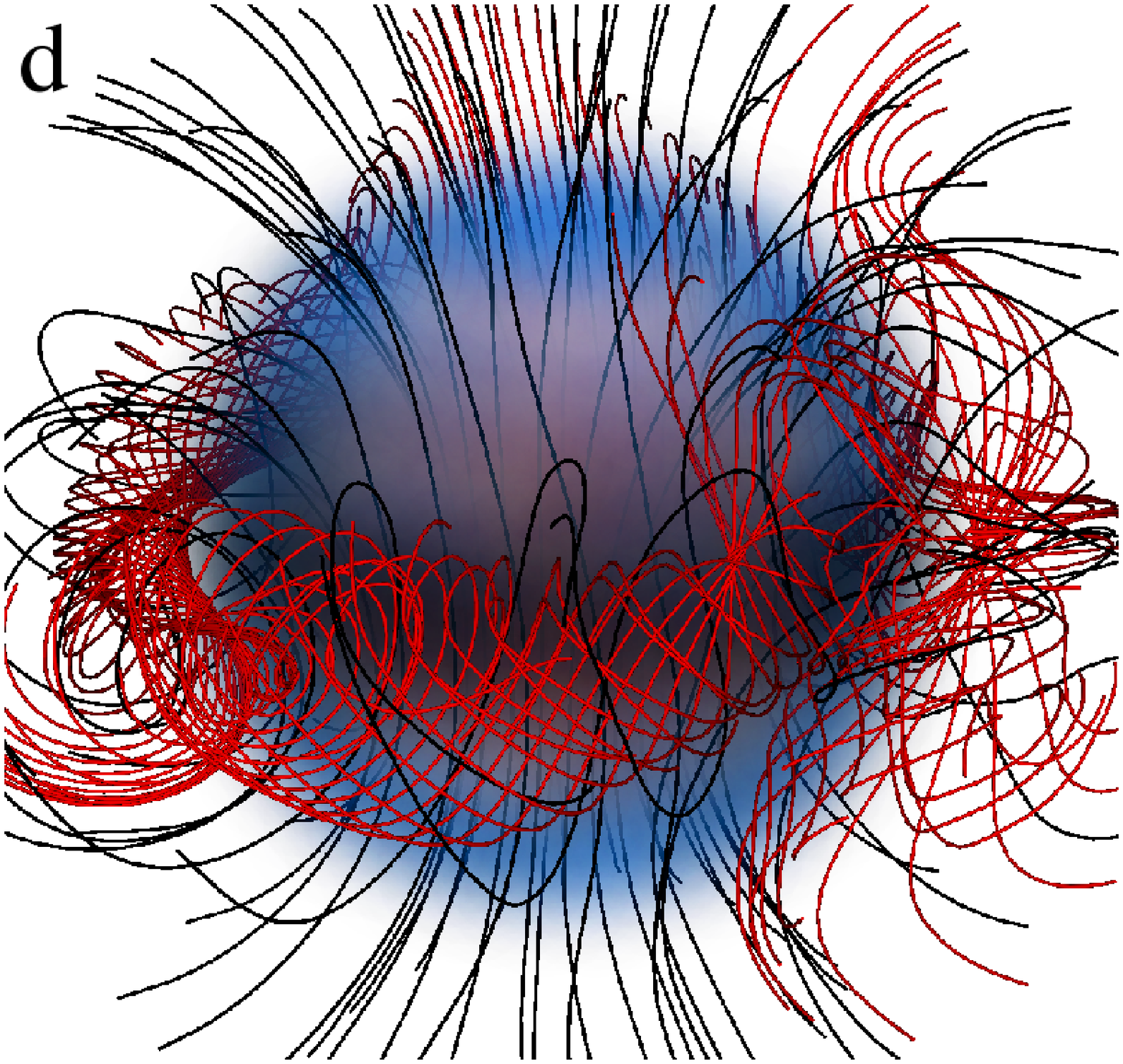}
	\caption{\label{fig2}  Time evolution of fiducial model with average magnetic field $\bar{B}=12.8\times10^{15}\,\mbox{G}$, corresponding to an Alfv\'en wave crossing time of $5\mbox{ms}$.  The figures are a) $t=0\,\mbox{ms}$, b) $t=25\,\mbox{ms}$, c) $t=50\,\mbox{ms}$ and d) $t=195\,\mbox{ms}$.  To more clearly visualise the instability, the red field lines are  seeded on the equatorial plane close to the neutral line, and the black field lines are seeded on the equatorial plane interior to the neutral line.  The volume rendering is an isopycnic surface at $37\%$ of the central rest-mass density, shown to provide contrast with the field lines.}
\end{figure}
\end{center}

In figures \ref{fig2} we present three-dimensional plots of the magnetic field at various instances throughout the evolution\footnote{A movie of the fiducial simulation lasting $400\,\mbox{ms}$ can be viewed at \url{http://www.tat.physik.uni-tuebingen.de/~tat/grmhd/}}.  The instability acts near the neutral line, which is the line where $B=0$ around the equatorial plane, located in our simulations at approximately two-thirds of the stellar radius.  For clarity, we have plotted red field lines seeded near the neutral line in the equatorial plane.  Also plotted are black field lines seeded on the equatorial plane interior to the neutral line.  The blue volume rendering is an isopycnic surface of $\rho=0.37\rho_{c}$, where $\rho_{c}$ is the central rest-mass density, which lies at a radius of approximately $50\%$ of the star; well inside the neutral line.  

Figure \ref{fig2}a shows the initial data imported from the {\sc lorene} spectral code.  The domain of our grid is larger than that plotted here and field lines are truncated at the surface of the star for clarity.  Such a configuration is known to be unstable to the kink instability from local linear studies in Newtonian physics \citep{markey73}.  Such studies have been confirmed with global, linearized, numerical evolutions \citep{lander11b} and in full non-linear studies \citep{braithwaite07}.  This is the first time such a configuration has been studied in general relativity.  

Figure \ref{fig2}b shows the evolution after $25\,\mbox{ms}=5\,t_{A}$.  The onset of the 'sausage' or 'varicose' mode \citep{markey73}, involving changes in the cross-sectional area of a flux tube around the neutral line, is clearly visible.  This is strongest in the $m=4$ mode, which is a result of the transient excitation at the beginning of our simulation.  Whilst this transient  reduces with increasing grid resolution, the presence of the varicose mode is an inherent characteristic of the system (as evidenced from its almost constant quasi-equilibra state that is resolution independent).  

The varicose mode is visually present in our simulation for approximately eight or nine Alfv\'en crossing times before the 'kink' instability appears and begins to dominate. In figure \ref{fig2}c we show the onset of this mode acting perpendicular to the gravitational field, in accordance with the prediction of \citet{markey73}.  This figure is after $50\,\mbox{ms}=10\,t_{A}$, at which point we can still see the presence of the varicose mode; i.e. the red field lines still exhibit varying cross-sectional area around the star.  It is worth noting that this mode will have been excited from the beginning of the simulation (as seen in figure \ref{fig1}), however it is only now that the exponential growth has reached a point at which it is visually obvious. In many ways this represents the non-linear development of the instability where the change in field structure is of similar order to the background field.  It is also worth mentioning here that our choice of the Cowling approximation is well justified considering the dominant instability acts on equipotential surfaces.

The modal analysis presented in figure \ref{fig1} does not distinguish between the varicose or kink modes, implying exponential growth of these quantities does not necessarily indicate an instability in one or the other mode.  Indeed, figures \ref{fig2}b and \ref{fig2}c suggest that, if the varicose mode is unstable it grows slower than the kink mode.  It is the kink mode that causes a cataclysmic reconfiguration of the topology of the magnetic field.  The linear analysis regarding the stability of the varicose mode is also unclear.  \citet{tayler57} showed the varicose mode is unstable in cylindrical geometry, however utilising a system that is not equivalent to a full neutron star.  \citet{lander11b} looked at the stability of poloidal fields in neutron stars by way of numerical evolutions of the linearized equations, however did not identify the difference between the varicose and kink modes.  Summarily, we believe the stability of the varicose mode to be an open question.

Figure \ref{fig2}d shows the simulation after $195\,\mbox{ms}=39\,t_{A}$.  This is a typical snapshot many Alfv\'en timescales after the non-linear saturation of the unstable modes.  We discuss the end-state in more detail in section \ref{equil}.

\subsection{Instability Growth-Times}
\begin{figure}
\epsscale{1.0}
	\plotone{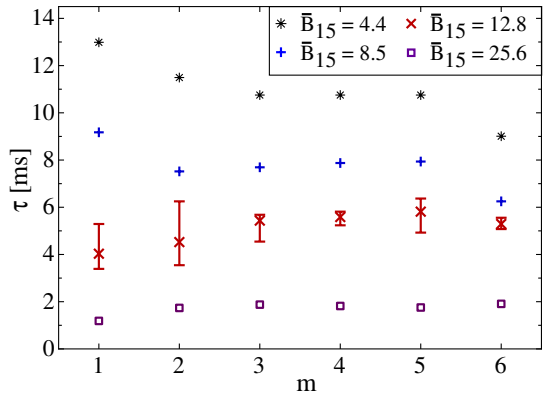}
	\caption{\label{fig3} Growth-time, $\tau$, as a function of mode number, $m$, for models with average magnetic field strengths ranging from $\bar{B}_{15}=4.4\,-\,25.6$, where $\bar{B}_{15}=\bar{B}/10^{15}\,\mbox{G}$.  We plot error bars on the fiducial, $\bar{B}_{15}=12.8$ simulation, which are derived from determining the growth-time scale, equation (\ref{growth}), over different regions of the instability.}
\end{figure}

We are interested in extracting instability growth-times for the individual modes.  We define the growth-time for each mode as
\begin{equation}
	\tau=\frac{\Delta t}{\Delta\ln\left[C_{m}\left(f\right)\right]},\label{growth}
\end{equation}
which we measure during the near exponential growth of the instability.  We have performed multiple simulations with different magnetic field strengths to examine the scaling behaviour of the instability.  In figure \ref{fig3} we plot the growth-times as a function of mode number for each model.  The scaling behaviour is consistent with what is anticipated, i.e. models with stronger magnetic fields and hence faster Alfv\'en velocities have faster growing instabilities.  This scales almost linearly; the model with average field strength $\bar{B}_{15}=4.4$ (where $\bar{B}_{15}=\bar{B}/10^{15}\,\mbox{G}$) has an Alfv\'en crossing time of $t_{A}=13.9\,\mbox{ms}$ and growth-time of approximately $\tau\sim12\,\mbox{ms}$, whereas the crossing time for the $\bar{B}_{15}=25.6$ model is $t_{A}=2.4\,\mbox{ms}$ and growth-time $\tau\sim2\,\mbox{ms}$.

Whilst this coarse measurement of growth-times scales correctly with magnetic field strength, for some of our models the individual modes do not scale in the manner necessarily expected.  \citet{kruskal54} and \citet{tayler57} showed the instability grows faster for increasing $m$ (our $m$ corresponds to their wavenumber $k$), akin to what is seen for our two simulations with the weakest magnetic fields, i.e. models with $\bar{B}_{15}=4.4$ and $8.5$ in figure \ref{fig3}.  This further agrees with the analysis of \citet{lander11b}, who showed growth-times converging soon after the $m=4$ mode.  We do note that our simulations with stronger magnetic fields show different behaviour in that the growth-time as a function of mode number is almost constant.

We provide three possibilities for the cause of the aforementioned discrepancy.  1) We are possibly seeing here mode coupling between the lower order modes, which may scale with magnetic field strength.  A more detailed analysis of this is required.  2) We are extracting a superposition of both the varicose and kink modes, which potentially have significantly different growth-times for the same mode number.  3) Our mode signal is 'dirty', in that when evaluating equation (\ref{growth}), we pick a region of the exponential growth and perform a linear regression (in log space) to determine the growth-time.  Choosing a different region of the exponential growth can give significantly varied growth-times.  This is displayed by way of error bars on the fiducial model in figure \ref{fig3}, which are of similar order to the size of variation one expects between the modes.  We expect this to be an inherent difficulty with extracting such growth timescales from non-linear simulations.  

\subsection{Equilibrium Configurations}\label{equil}
We now return to figure \ref{fig2}d, particularly concerning our equilibrium configurations.  In some ways, figure \ref{fig2}d is consistent with the `twisted-torus' configurations seen in the non-linear evolutions of \citet{braithwaite06b} and \citet{braithwaite09}, and in the semi-analytic equilibrium derivations of \citet{yoshida06b} and \citet{ciolfi09}.  This figure shows about half of the star is well approximated by a twisted-torus, however the remainder of the star also exhibits large non-axisymmetric structures.  This is typical of the evolution following the saturation of the instability.  Whilst \citet{braithwaite08} has also found non-axisymmetric equilibria as a result of various non-linear simulations, we believe ours are the first such simulations to do so with a barotropic equation of state.  This therefore extends existing results in the literature \citep[e.g.][]{ciolfi09}, in that we provide evidence for a new branch of stable, equilibrium solutions with a barotropic equation of state.  

A way of classifying variation from the initial state is through calculating the relative energy in the toroidal and poloidal components of the magnetic field.  The initial field obviously contains $E_{p}/E=1.0$, where $E_{p}$ and $E$ are respectively the poloidal and total magnetic field energy's.  Throughout the evolution of the instability we see a transfer of energy between the toroidal and poloidal energies, although it is interesting to note that this does not take place until we visually see the onset of the kink instability (i.e. $\sim50\,\mbox{ms}$).  Our equilibrium configuration is then characterised by a magnetic field with $E_{p}/E\sim0.75$, which is similar to the upper-end of the stability window found in stably stratified models of \citet{braithwaite09}.  

It can further be seen from figure \ref{fig2}d and the full evolution of the system that, despite large-scale restructuring of the interior field, the center of the star is still threaded by a dominantly poloidal component.  Moreover, the magnetic field exterior to the star is also dominantly poloidal, with little visual change to its initial structure.  This is also important for stability issues associated with poloidal fields, particularly the \citet{flowers77} instability which relies on minimising the energy in the exterior region of the star \citep[see recent result of][]{marchant10}.  

\section{Conclusion}
We have presented the first, non-linear GRMHD simulations of hydromagnetic instabilities inherent to purely poloidal magnetic fields.  In particular, we have evolved initially self-consistent solutions of the Einstein-Maxwell field equations under the assumption of ideal MHD and the Cowling approximation.  Purely poloidal magnetic field configurations are particularly vulnerable to two modes -- the varicose/sausage and kink modes.  The varicose mode, a cross-sectional change in a toroidal tube of magnetic field around the neutral line, was initially excited in our simulations due to the Cartesian grid.  This mode evolved to a pseudo-equilibrium (figure \ref{fig2}b) before it was swamped by the kink instability (figure \ref{fig2}c), which acts perpendicular to the gravitational field.  The kink instability saturated after approximately 15 Alfv\'en crossing times, at which point the system evolved towards a new, non-axisymmetric equilibrium (figure \ref{fig2}d).  

The equilibrium end-states achieved from our simulations partially resemble twisted-torus configurations seen from other non-linear evolutions \citep{braithwaite06b,braithwaite08}.  The center of the star is threaded by a dominantly poloidal component, which is consistent with the instability dominating near the neutral line of the star.  In some regions of the star we see the presence of twisted flux tubes, with other regions displaying highly non-axisymmetric portions of the magnetic field.  These are the first non-axisymmetric equilibria attained with a barotropic equation of state.  An interesting extension to this work is to study the evolution of the exterior magnetic field as a result of the interior rearrangement.  Observations of the spin down of neutron stars only infers the dipole component of the field, implying the presence of extra degrees of freedom can only be inferred at this stage through detailed modeling.  However, a better understanding of the dynamics of the stellar surface in our simulations is required to accomplish this goal.  

There are many further extensions of this work worth pursuing towards more realistic neutron star models.  Such models should feature composition stratification, which is likely to have an impact on MHD equilibrium configurations \citep[e.g.][]{reisenegger09}.  Further improvements can include crustal \& magnetospheric physics and studying the effects of superfluidity/superconductivity \citep[e.g.][]{glampedakis11}.  

The end-state quasi-equilibria models discovered herein add to the possible solution space of barotropic equilibria.  Axisymmetric equilibrium models have been explored by \citet{ciolfi09,ciolfi10}, which would be interesting to test for stability.  Indeed, any pseudo-stable equilibrium could serve as possible starting points for studying the secular magnetic field evolution in magnetars and its connection to activity seen in these objects.  Moreover, the theoretical modeling of magnetar QPOs (which has so far been done using symmetric magnetic field configurations), should be extended to account for the non-axisymmetric configurations akin to those encountered here.  We hope to address some of these exciting issues in the near future.

\acknowledgments
This work is supported by the Transregio 7 `Gravitational Wave Astronomy', financed by the Deutsche Forschungsgemeinschaft (DFG).  PL and KG are supported by the Alexander von Humboldt Foundation.  We are grateful to Jerome Novak and Erik Schnetter for assistance with {\tt Lorene}, and particularly grateful to Jon Braithwaite for useful discussions.


\end{document}